\documentclass[manuscript,screen]{acmart}

\usepackage{multirow}

\AtBeginDocument{%
  \providecommand\BibTeX{{%
    \normalfont B\kern-0.5em{\scshape i\kern-0.25em b}\kern-0.8em\TeX}}}

\setcopyright{acmcopyright}
\copyrightyear{2021}
\acmYear{2021}
\acmDOI{10.1145/}

\acmConference[RecSys '21]{RecSys '21: ACM Conference on Recommender Systems}{September 27 -- October 1, 2021}{Amsterdam, the Netherlands}
\acmBooktitle{RecSys '21: ACM Conference on Recommender Systems,
  September 27 -- October 1, 2021, Amsterdam, the Netherlands}
\acmPrice{15.00}
\acmISBN{978-1-4503-XXXX-X/18/06}

\begin{document}

\title{Sequence Adaptation via Reinforcement Learning in Recommender Systems}

\author{Stefanos Antaris}
\email{antaris@kth.se}
\affiliation{
 \institution{KTH Royal Institute of Technology, Hive Streaming AB}
 \country{Sweden}
}
\author{Dimitrios Rafailidis}
\email{draf@uth.gr}
\affiliation{%
 \institution{University of Thessaly}
 \country{Greece}
}

\renewcommand{\shortauthors}{}

\begin{abstract}

Accounting for the fact that users have different sequential patterns, the main drawback of state-of-the-art recommendation strategies is that a fixed sequence length of user-item interactions is required as input to train the models. This might limit the recommendation accuracy, as in practice users follow different trends on the sequential recommendations. Hence, baseline strategies might ignore important sequential interactions or add noise to the models with redundant interactions, depending on the variety of users' sequential behaviours. To overcome this problem, in this study we propose the SAR model, which not only learns the sequential patterns but also adjusts the sequence length of user-item interactions in a personalized manner. We first design an actor-critic framework, where the RL agent tries to compute the optimal sequence length as an action, given the user's state representation at a certain time step. In addition, we optimize a joint loss function to align the accuracy of the sequential recommendations with the expected cumulative rewards of the critic network, while at the same time we adapt the sequence length with the actor network in a personalized manner. Our experimental evaluation on four real-world datasets demonstrates the superiority of our proposed model over several baseline approaches. Finally, we make our implementation publicly available at \url{https://github.com/stefanosantaris/sar}.

\end{abstract}




\maketitle

\section{Introduction}

Recommender systems are essential for several real-world applications such as online services~\cite{Hansen2020Recsys,Chen2019fashion}, point of interest platforms~\cite{ferraro2020exploring,Sun_Qian_Chen_Liang_Nguyen_Yin_2020,Lim2020cikm}, social networks~\cite{Song2019wsdm,Qin2020wsdm}, and so on. The goal of sequential recommender systems is to capture the sequential pattern of a user to generate accurate next item recommendations. Towards this aim, significant progress has been made in personalized sequential recommendation methods, focusing on modelling the ordered sequence of user-item interactions~\cite{hidasi2016sessionbased,Hidasi_2018,Ji2020MTAM,Wang2019sasrec,wu2020ssept,Xin202sassqn}. Baseline strategies design recurrent neural networks to capture personalized preferences~\cite{hidasi2016sessionbased, Hidasi_2018}, while other sequential recommendation models adopt attention mechanisms~\cite{Ji2020MTAM} and transformers~\cite{Wang2019sasrec, wu2020ssept}. Recently, Reinforcement Learning (RL) strategies have been proposed to model sequential preferences by designing an agent to interact with users~\cite{Hu2018RLSRS,LEI2020AttentiveQNetwork}. The key idea is to train a RL agent to model personalized preferences based on the sequential user-item interactions and produce recommendations as actions accordingly. For instance, Xin et al.~\cite{Xin202sassqn} present a self-supervised framework to consider the values of the RL agent as regularization to the sequential prediction algorithm. 

A downside of baseline sequential recommendation models is that they set a fixed sequence length, which means that a predefined number of past user-item interactions is required as input to train a model and generate sequential recommendations. By considering a fixed sequence length, the baseline models fail short to accurately model users' preferences in a personalized manner. This happens because each user does not only have different sequential preferences, but also different tendencies to change her preferences over time. This means that an adaptive way is required to compute the sequence length of the user-item interactions in a personalized manner. For example, a user might be interested in buying laptop accessories immediately after buying a laptop, while another user completely omits the laptop accessories and the next item in the sequence might be a book. By applying a fixed sequence length, the accuracy of the sequential recommendation models is limited.  Therefore, it is crucial for the recommendation models to adjust the length of the input sequence for each user. Recently, adaptive sequence submodularity techniques have been exploited to adapt the sequence length of user-item interactions for recommender systems~\cite{Tschiatschek2017submodularity}. For instance, instead of considering a fixed sequence length, Mitrovic et al.~\cite{Mitrovic2019adapt} first model the sequential user-item interactions as a directed acyclic graph and then an adaptive sequence greedy policy is followed to select a subset of items based on the $k$-nearest user-item interactions. However, this strategy ignores the order of the previous user interactions in the subset of the $k$-nearest user-item interactions, thus the model underperforms in capturing users' sequential patterns. 

To overcome the shortcomings of baseline strategies, in this paper we propose a \textbf{S}equence \textbf{A}daptation model via deep \textbf{R}einforcement learning in recommender systems, namely SAR, making the following contributions:
\textbf{(i)} \emph{To calculate the sequence length in an adaptive way for each user, we formulate the selection of the interactions' sequence length as a Markov Decision Process (MDP) and follow a deep RL strategy. In particular, the RL agent training considers the selection of a sequence length as an action in an actor-critic framework. In doing so, we learn a global policy and generate an adaptive sequence of interactions for each user at a certain time step.} \textbf{(ii)} \emph{Then, we design a joint loss function, where the RL agent with the adapted sequence of interactions acts as a regularizer to produce the recommendations}. Our experiments with four benchmark datasets demonstrate that SAR outperforms several state-of-the-art strategies. The remainder of the paper is organized as follows, in Section~\ref{sec:model} we formally define the problem of adaptive sequential recommendations and detail the SAR model. Our experimental evaluation is presented in Section~\ref{sec:exp}, and we conclude our study in Section~\ref{sec:conc}.

\section{Proposed Model} \label{sec:model}

\subsection{Problem Formulation}

Let $\mathcal{U}$ and $\mathcal{I}$ be the set of users and items, respectively. For each user $u \in \mathcal{U}$ we consider an ordered sequence $x_{ut} = \{i_1, \ldots, i_t\}$, where $i_t \in \mathcal{I}$ is the item that the user $u$ has interacted at the time step $t$. The goal of our model is to recommend an item $i_{t+1}$, that is the most relevant item to the user's $u$ preferences based on the adapted sequence $x'_{ut}= \{i_{t-l}, \ldots, i_{t}\}$, with $l \leq t$ being the number of the last interacted items. Note that compared to state-of-the-art strategies, instead of fixing the sequence length, the number of the last interacted items $l$ in our model has also to be adjusted for each user $u$ at the $t$-th step.

We model the selection of the $l$ previous items in the sequence $x_{ut}$ of user $u$ as a Markov Decision Process (MDP). A MDP is defined as $\mathcal{M} = (\mathcal{S}, \mathcal{A}, \mathcal{P}, \mathcal{R}, \gamma)$, with $\mathcal{S}$, $\mathcal{A}$, $\mathcal{P}$, $\mathcal{R}$ and $\gamma$ being the state, action, transition probability, reward sets and the discount factor, respectively. At each time step $t$, the agent takes an action $a_t \in \mathcal{A}$ to select the $l \leq t$ last items based on the state $s_t \in \mathcal{S}$ of user $u$. Given the action $a_t$, we generate the sequence $x'_{ut}$ with the  $l$ latest interactions, to recommend the next item $i_{t+1}$ for a user $u$ at the $t$-th time step. The agent receives the accuracy of the recommended item $i_{t+1}$ as a reward $R_t(a_t,s_t) \in \mathcal{R}$ for the selected action $a_t$ and user $u$ transitions to state $s_{t+1}$ with probability $p(s_{t+1}| s_{t}, a_t) \in \mathcal{P}$. The goal of the agent is to find $\forall$ $u \in \mathcal{U}$ the optimal sequence length $l$, so as to maximize the sequential recommendation accuracy. In our model, the agent optimizes the Bellman equation by maximizing the state-action transition value $Q(s_t,a_t) = \max_{\pi_{\theta}} \mathbb{E}_{\pi_{\theta}} \{ \sum_{k=0}^t \gamma^k R_k(a_k, s_k)\}$, where $\mathbb{E}_{\pi_{\theta}}$ is the expectation based on the policy $\pi_{\theta}:\mathcal{S} \times \mathcal{A} \rightarrow [0,1]$ and $\theta$ are the parameters of the policy $\pi$.

\subsection{Overview of SAR}

\begin{figure}
    \centering
    \includegraphics[scale=0.42]{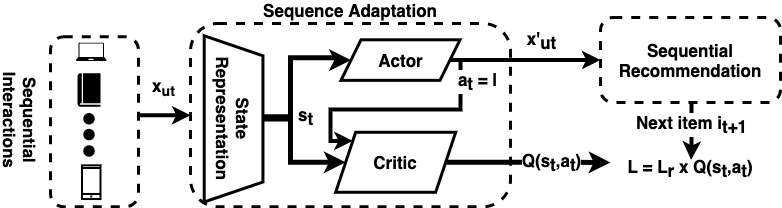}
    \caption{Overview of SAR. The Sequence Adaptation component generates the adapted sequence $x'_{ut}$ with the $l$ latest interactions via an actor-critic framework. Then, the adapted sequence $x'_{ut}$ is provided as input to the Sequential Recommendation component to produce the final recommendations by optimizing the joint loss function $L$.}
    \label{fig:overview}
\end{figure}

As illustrated in Figure \ref{fig:overview}, the SAR model consists of two main components: (i) the \emph{Sequence Adaptation} and (ii) \emph{Sequential Recommendation} components. The goal of the proposed SAR model is to first adjust the length $l$ and generate the adapted sequence $x'_{ut}$ for each user $u$, and then to recommend the next item $i_{t+1}$. \emph{Sequence Adaptation:} The role of this component is to learn a global policy $\pi_{\theta}$ that calculates the length $l$ of the sequence $x'_{ut}$ for each user $u$ at the time step $t$. The input of the Sequence Adaptation component is the sequence $x_{ut} = \{i_{1}, \ldots, i_t\}$, which consists of all the items $i_t \in \mathcal{I}$ that the user $u$ interacted with up to the time step $t$. We design an actor-critic reinforcement learning scheme to adapt the sequence length $l$ for each user $u$. The Sequence Adaptation component outputs the adapted sequence $x'_{ut} =\{i_{t-l}, \ldots, i_{t}\}$, which contains the $l$ latest interactions of user $u$. \emph{Sequential Recommendation:} The goal of this component is to recommend the next item $i_{t+1}$ for a user $u$. Provided the adapted sequence $x'_{ut}$ as input, in our implementation we design a personalized transformer model, regularized by the actor-critic framework to generate the sequential recommendations by optimizing the joint loss function $L$.

\subsection{Sequence Adaptation Component}

The input of the Sequence Adaptation component is the sequence $x_{ut} = \{i_1, \ldots, i_{t}\}$, which consists of all the interacted items by user $u$ until the time step $t$. Following an actor-critic  scheme~\cite{konda2000actor}, we learn the optimal sequence length $l$ based on the state $s_t$ for each user $u$. The Sequence Adaptation component consists of three main parts: i) the State Representation, ii) the Actor Network, and iii) the Critic Network.

\textbf{State Representation.} At each time step $t$ the state representation takes as input the sequence $x_{ut} = \{i_1, \ldots, i_t\}$ with all the interacted items up to the time step $t$, and outputs a $d_s$-dimensional state representation vector $\mathbf{s}_t \in \mathbb{R}^{d_s}$ which corresponds to the state $s_t$ of user $u$. We represent each item $i_t$ of the sequence $x_{ut}$ as a $d_i$-dimensional vector $\mathbf{i}_t \in \mathbb{R}^{d_i}$ and each user $u\in\mathcal{U}$ as a $d_u$-dimensional vector $\mathbf{u} \in \mathbb{R}^{d_u}$. Given the sequence $x_{ut}$, we compute the $d_s$-dimensional state vector $\mathbf{s}_t$ as follows~\cite{Parisotto2020gtrxl}:

\begin{equation}
    \mathbf{s}_t = GTrXL(\mathbf{x}_{ut})
\end{equation}
where GTrXL is the Gated Transformer-XL which provides robustness when applying transformers on RL agents~\cite{vaswani2017attention}. Note that GTrXL consists of attention layers to compute the similarity among consecutive items. This means that the state representation vector $\mathbf{s}_t$ captures the evolution of users' preferences. 

\textbf{Actor Network.} The actor network takes as input the state vector $\mathbf{s}_t$ and then computes the action $a_t=l \in 1,\ldots,t$, which in practice is the selection of the sequence length $l$, with $1 \leq l \leq t$. Given that users' preferences change over time, at the $t$-th step the actor network leverages the state vector $\mathbf{s}_t$ to identify the $l$ latest interactions. In particular, for each time step $t$ the actor network transforms the state vector $\mathbf{s}_t$ of user $u$ by employing a two-layer perceptron (MLP):

\begin{equation} \label{eq:act}
    a_t = f\bigg( \sigma(\mathbf{s}^{\top}_t\mathbf{W}^a_1 )  \mathbf{W}^a_2\bigg)
\end{equation}
where $\mathbf{W}^a_1 \in \mathbb{R}^{d_s \times d_{a}}$ and $\mathbf{W}^a_2 \in \mathbb{R}^{d_a \times 1}$ are the parameter weights, and $ d_{a}$ is the number of dimensions of the MLP's intermediate layer. The symbols $f$ and $\sigma$ are the ReLU and sigmoid activation functions, respectively. This means that for each time step $t$ the sequence length $l$ of user $u$ corresponds to the action $ a_t $, calculated in Equation~\ref{eq:act}. During the model training (Section~\ref{sec:src}) the adapted sequence $x'_{ut}$ is generated by selecting the $l$ last items of the sequence $x_{ut}$.

\textbf{Critic Network.} The input of the critic network is the state vector $\mathbf{s}_t$ and the action $a_t$ generated by the actor network. The critic network outputs a scalar value $Q(\mathbf{s}_t,a_t) \in \mathbb{R}^{+}$, which corresponds to the expected reward of the agent's action $a_t$, given the user's state $\mathbf{s}_t$. We compute the value $Q(\mathbf{s}_t,a_t)$ by employing a two-layer MLP as follows:

\begin{equation}
    Q(\mathbf{s}_t, a_t) = f\bigg( \sigma((\mathbf{s}^{\top}_t \oplus a_t)\mathbf{W}^c_1 )  \mathbf{W}^c_2\bigg)
    \label{eq:qvalues}
\end{equation}
where $\mathbf{W}^c_1 \in \mathbb{R}^{d_s + 1 \times d_{c}}$ and $\mathbf{W}^ac_2 \in \mathbb{R}^{d_c \times 1}$ are the parameter weights, and $ d_{c}$ is the number of dimensions of the MLP's intermediate layer. The symbol $\oplus$ is the concatenation operation of the state vector $\mathbf{s}_t$ with the action $a_t$.

% \textbf{Sequence Generation} As shown in Figure~\ref{fig:overview}, the input of sequence generation is the action $a_t$, that is the sequence length $l$, with $0 < l \leq t$, computed by the actor network, and the sequence interactions $x_{ut}$ of user $u$ with all the interactions until the $t$-th step. During training of the actor-critic network, the actor network computes the optimal sequence length $l$ for each user based on the action $a_t$ in Equation~\ref{eq:act}. 

% Given the sequence length $l$. To compute the optimal sequence length $l$, we $\forall l=1,ldots,t$ 

% Provided that the state representation component captures the user preferences, the sequence generation component requires no trainable parameters to generate the adapted sequence $x'_{ut}$. Therefore, the output of the sequence generation component is computed by extracting the $l$ latest user interactions from the input $x_{ut}$.

\subsection{Sequential Recommendation Component} \label{sec:src}

The Sequential Recommendation component takes as an input the sequence vector $x'_{ut}$ generated by the Sequence Adaptation component. To compute the recommended item $i_{t+1}$, we implement the following transformer architecture~\cite{wu2020ssept}:

\begin{equation}
% \begin{array}{c}
    i_{t+1} = argmax(\sigma(\mathbf{F}_t \cdot (x'_{ut} \oplus \mathbf{u})))  
    % \mathbf{r} =  
% \end{array}
\end{equation}
where $\mathbf{u}$ is the $d_u$-dimensional vector representation of user $u$, and $\mathbf{F}_t \in \mathbb{R}^{|I| \times d_u}$ is the set of the output hidden units produced by the transformer's encoder at the $t$-th time step.

In the Sequential Recommendation component, we define the cross-entropy function as follows~\cite{Xin202sassqn, wu2020ssept}:

\begin{equation} \label{eq:cross}
    L_r = - \sum_{k=1}^{|I|} Y_k log(r_k)
\end{equation}
where $Y_k = 1$, if the user interacted with the item $i_k$ at the $t+1$ timestamp and 0, otherwise. 

Given that the input sequence $x'_{ut}$ of the Sequential Recommendation component is parameterized by the action $a_t$ of the Sequence Adaptation component, we use the Q-values (Equation \ref{eq:qvalues}) to regularize the cross-entropy function in Equation~\ref{eq:cross}. Therefore, the SAR model is trained by optimizing the following joint loss function:

\begin{equation}
    L = L_r \cdot Q(\mathbf{s}_t,a_t) \label{eq:loss}
\end{equation}

\textbf{Model Training.} Overall, we train our SAR model for several epochs by optimizing the joint loss function in Equation \ref{eq:loss}. In particular, we optimize the trainable parameters of both the Sequence Adaptation and Sequential Recommendation components. In doing so, we align the  accuracy of the Sequential Recommendation component with the expected cumulative reward of the critic network, while at the same time we adapt the sequence length with the actor network in a personalized manner. 

\begin{table}[t]
    \centering
    \caption{Dataset statistics}
    \begin{tabular}{c|c|c|c}
        \hline
        \textbf{Dataset} & \textbf{\#Users} & \textbf{\#Items} & \textbf{Avg. Seq. Length}  \\ \hline
        \textbf{Steam} & $332,924$ & $12,773$ & $11.05$  \\ \hline
        \textbf{ML-10M} & $69,877$ & $10,436$ & $143.10$ \\ \hline
        \textbf{Electronics} & $650,866$ & $142,064$ & $5.13$ \\ \hline
        \textbf{Kindle} & $280,658$ & $126,717$ & $7.55$  \\ \hline
    \end{tabular}
    \label{tab:datasets}
\end{table}

\section{Experimental Evaluation} \label{sec:exp}

\textbf{Datasets} In our experiments, we evaluate our proposed SAR model on four real-world datasets. The Steam dataset contains users' reviews on a video game platform~\cite{Wang2019sasrec}. The ML-10M\footnote{\url{https://grouplens.org/datasets/movielens/10m/}} dataset contains 10 million movie ratings of the movielens platform. Electronics and Kindle\footnote{\url{https://jmcauley.ucsd.edu/data/amazon/}} datasets contain reviews of Amazon products. In Table \ref{tab:datasets}, we summarize the statistics of each dataset and report the average sequential length of the user-item interactions.

\textbf{Evaluation Protocol} We evaluate the performance of our proposed model in terms of two ranking metrics Normalized Discount Cumulative Gain (NDCG) and Hit Ratio (HR), which are defined as follows: 
\begin{equation}
\begin{array}{cc}
    NDCG@K = \frac{1}{|\mathcal{U}|} \sum_{u \in \mathcal{U}} \frac{DCG@K(u, \mathbf{r})}{DCG@K(u,\mathbf{r}^{*})},  & \text{with} \indent
    
    DCG@K(u, \mathbf{r}) = \sum_{i = 1}^{K} \frac{2^{r_j} - 1}{log_2 (j + 1)} \\ \\
    
    \multicolumn{2}{c}{HR@K = \frac{1}{|\mathcal{U}|} \sum_{u \in \mathcal{U}} \sum_{k=1}^{K} Y_k}
\end{array}
\end{equation}
where $\mathbf{r}$ corresponds to the predicted probabilities of the items in the set $\mathcal{I}$ and $\mathbf{r}^{*}$ is the ground-truth probabilities. The term $Y_k$ indicates if a user $u$ interacts with each of the $K$ recommendations. The NDCG metric measures the relative position of the predicted item in the $K$ recommendations and the HR metric is a recall-based metric which measures how many ground-truth items have been predicted in the $K$ recommendations. Following the evaluation protocol of~\cite{wu2020ssept}, we train each examined model with all the interacted items of each user up to the time step $t-2$. In our experiments, for each user we consider the second last interacted item as validation and the last item as testing. We repeated our experiments five times and report average NDCG and HR values. 

\textbf{Compared Methods} We evaluate the performance of the proposed SAR model against the following sequential recommendation strategies: i) SSE-PT++\footnote{\url{https://github.com/wuliwei9278/SSE-PT}}~\cite{wu2020ssept} is a sequential recommendation model that employs personalized transformers; ii) SASRec-SAC\footnote{\url{https://drive.google.com/open?id=1nLL3_knhj_ RbaP_IepBLkwaT6zNIeD5z}}~\cite{Xin202sassqn} is a RL-based approach that exploits the actor-critic scheme to regularize the recommendation strategy; iii) SASRec-SQN is a variant of SASRec-SAC that employs a Deep Q-Network instead of the actor-critic scheme; iv) MTAM\footnote{\url{https://github.com/cocoandpudding/MTAMRecommender}}\cite{Ji2020MTAM} is a time-aware sequential recommendation strategy that designs attentive memory networks to model the time difference among consecutive user-item interactions; v) TiSASRec\footnote{\url{https://github.com/JiachengLi1995/TiSASRec}}~\cite{Li2020TISASREC} is a baseline approach that adopts the positional attention mechanism to capture the evolution of users' preferences over time; and vi) AdaptiveSubSeq\footnote{\url{https://github.com/ehsankazemi/adaptiveSubseq}}~\cite{Mitrovic2019adapt} is a sequence adaptation recommendation strategy that exploits submodularity techniques to approximate the recommendations by identifying the $k$ nearest user-item interactions in a directed acyclic graph.

\textbf{Configuration} For each examined model, we used the publicly available implementations and tuned the hyperparameters following a grid-selection strategy on the validation set. In SSE-PT++ we set the dimensions of user and item embeddings to $100$ and the sequence length to $l=50$. In SASRec-SAC and SASRec-SQN, we fix the user and item embeddings to $200$, the state representation vector $\mathbf{s}$ of a user to $100$ and the discount factor $\gamma$ to $0.94$. Moreover, the sequence length of SASRec-SAC and SASRec-SQN is set to $l=100$. MTAM uses $100$-dimensional user and item embeddings and the sequence length is fixed to $l=30$. In TiSASRec the user and item embeddings are set to $150$ and the sequence length to $100$. AdaptiveSubSeq models the user-item interactions as a directed acyclic graph, thus user and item embeddings are not considered. Given that the sequence length is adaptive, we fix the maximum allowed sequence length to $l=200$. In the proposed SAR model, we set the user and item embeddings to $100$, the state representation vector $\mathbf{s}_t$ is fixed to $150$ dimensions and the discount factor $\gamma$ is set to $0.82$. We train each model for $100$ epochs and employ mini-batch gradient decent with Adam optimizer to compute the model parameters \cite{kingma2017adam}. 

In addition, to train the RL based approaches, that is SASRec-SAC, SASRec-SQN and SAR, we employ an offline batch RL environment~\cite{kumar2020offline,Singh2020cog,Fujimoto2019icml}. This means that the agent receives the sequential user interactions, ordered based on the interaction timestamp. To optimize the model parameters for each RL strategy, the agent is trained in an episodic manner~\cite{sutton2018reinforcement}. For each episode the agent receives the user's $u$ sequential interactions $x_{ut}$ and predicts the next item $i_{t+1}$~\cite{Xin202sassqn}. The episode is completed when the agent has received all the users' interactions. All our experiments were conducted on a single server with an Intel Xeon Bronze 3106, 1.70GHz CPU, with training acceleration by the GPU Geforce RTX 2080 Ti graphic card.

\begin{table}[t]
    \centering
    \caption{Methods' comparison, where bold values indicate the best method using a statistical significance t-test with $p < 0.05$.}
    \resizebox{\textwidth}{!}{
    \begin{tabular}{c|c|c|c|c|c|c|c|c}
        \hline
        \textbf{Dataset}        & \multicolumn{4}{c|}{\textbf{Steam}} & \multicolumn{4}{c}{\textbf{ML-10M}} \\ \hline
                                & \textbf{NDCG@10}  & \textbf{HR@10}    & \textbf{NDCG@5} & \textbf{HR@5}   & \textbf{NDCG@10}  & \textbf{HR@10}    & \textbf{NDCG@5}   & \textbf{HR@5} \\ \hline
        \textbf{SSE-PT++}       & $.523 \pm .024$   & $.323 \pm .033$   & $.462 \pm .021$ & $.304 \pm .058$ & $.455 \pm .052$   & $.209 \pm .022$   & $.424 \pm .014$   & $.133 \pm .062$  \\ \hline 
        \textbf{SASRec-SAC}     & $.543 \pm .032$   & $.334 \pm .011$   & $.526 \pm .012$ & $.311 \pm .024$ & $.482 \pm .033$   & $.252 \pm .034$   & $.441 \pm .041$   & $.192 \pm .053$ \\ \hline
        \textbf{SASRec-SQN}     & $.562 \pm .013$ & $.342 \pm .042$ & $.523 \pm .016$ & $.314 \pm .051$ & $.462 \pm .084$ & $.250 \pm .093$ & $.428 \pm .016$ & $.188 \pm .053$ \\ \hline
        \textbf{MTAM}           & $.512 \pm .054$   & $.309 \pm .022$   & $.406 \pm .063$ & $.301 \pm .033$ & $.426 \pm .013$   & $.227 \pm .043$   & $.401 \pm .022$ & $.173 \pm .042$ \\ \hline 
        \textbf{TiSASRec}       & $.507 \pm .023$   & $.332 \pm .024$   & $.477 \pm .082$ & $.284 \pm .062$ & $.484 \pm .049$   & $.233 \pm .034$   & $.432 \pm .046$ & $.154 \pm .068$\\ \hline
        \textbf{AdaptSubSeq}    & $.588 \pm .052$   & $.353 \pm .011$   & $.548 \pm .042$ & $.306 \pm .054$ & $.517 \pm .033$   & $.282 \pm .052$   & $.455 \pm .061$ & $.203 \pm .036$ \\ \hline
        \textbf{SAR}    & \textbf{.623} $\pm$ \textbf{.032}   & \textbf{.382} $\pm$ \textbf{.033}  & \textbf{.568} $\pm$ \textbf{.092} & \textbf{.332} $\pm$ \textbf{.072} & \textbf{.553} $\pm$ \textbf{.022}   & \textbf{.303} $\pm$ \textbf{.022}   & \textbf{.534} $\pm$ \textbf{.062} & \textbf{.224} $\pm$ \textbf{.026} \\ \hline \hline
        \textbf{Dataset}        & \multicolumn{4}{c|}{\textbf{Electronics}} & \multicolumn{4}{c}{\textbf{Kindle}} \\ \hline
        & \textbf{NDCG@10}      & \textbf{HR@10}    & \textbf{NDCG@5}   & \textbf{HR@5} & \textbf{NDCG@10} & \textbf{HR@10} & \textbf{NDCG@5}   & \textbf{HR@5} \\ \hline
        \textbf{SSE-PT++}       & $.462 \pm .042$   & $.222 \pm .024$   & $.343 \pm .041$ & $.150 \pm .022$ & $.623 \pm .052$ & $.354\pm .031$  & $.584 \pm .082$ & $.301 
        \pm 0.05$ \\ \hline
        \textbf{SASRec-SAC}     & $.488 \pm .043$ & $.236 \pm .051$ & $.415 \pm .063$ & $.204 \pm .052$ & $.677 \pm .032$ & $.374 \pm .042$ & $.604 \pm .092$ & $.312 \pm .083$ \\ \hline 
        \textbf{SASRec-SQN}     & $.482 \pm .043$ & $.231 \pm .023$ & $.402 \pm .033$ & $.189 \pm .019$ & $.682 \pm .041$ & $.383 \pm .073$ & $.612 \pm .052$ & $.326 \pm .066$ \\ \hline 
        \textbf{MTAM}           & $.433 \pm .032$ & $.211 \pm .034$ & $.383 \pm .052$ & $.174 \pm .021$ & $.613 \pm .046$ & $.328 \pm .052$ & $.566 \pm .062$ & $.318 \pm .042$ \\ \hline
        
        \textbf{TiSASRec}       & $.483 \pm .031$ & $.211 \pm .032$ & $.442 \pm .063$ & $.183 \pm .061$ & $.603 \pm .054$ & $.322 \pm .021$ & $.554 \pm .032$ & $.283 \pm .046$\\ \hline
        \textbf{AdaptSubSeq}    & $.522 \pm .023$ & $.263 \pm .022$ & $.504 \pm .062$ & $.235 \pm .052$ & $.723 \pm .022$ & $.393 \pm .054$ & $.617 \pm .082$ & $.326 \pm .063$\\ \hline
        \textbf{SAR}    & \textbf{.554} $\pm$ \textbf{.024} & \textbf{.332} $\pm$ \textbf{.041} & \textbf{.512} $\pm$ \textbf{.079} & \textbf{.303} $\pm$ \textbf{.048} & \textbf{.744} $\pm$ \textbf{.033} & \textbf{.423} $\pm$ \textbf{.041} & \textbf{.638} $\pm$ \textbf{.043} & \textbf{.404} $\pm$ \textbf{.063}\\ \hline
    \end{tabular}
    }
    \label{tab:ndcg}
\end{table}

\subsection{Performance Evaluation}
In Table~\ref{tab:ndcg}, we evaluate the performance of the examined models in terms of NDCG and HR. The proposed SAR model constantly outperforms the baseline approaches in all datasets. This suggests that SAR effectively captures the sequential patterns and provides accurate recommendations. Compared with the second best method AdaptSubSeq, the proposed SAR model achieves relative improvements $6.5$ and $8.4\%$ in terms of NDCG and HR in Steam, $12.1$ and  $8.6\%$ in ML-10M, $3.9$ and $27.6\%$ in Electronics, $3.2$ and $15.8\%$ in Kindle. We can also observe that AdaptSubSeq outperforms all the other sequential recommendation baselines, as AdaptSubSeq also computes adaptive sequences to generate personalized recommendations. This indicates the importance of the sequence adaptation in recommender systems. However, AdaptSubSeq ignores the order of the previous user interactions when computing the subset of the $k$-nearest user-item interactions in the acyclic graph. To solve this problem, in the proposed SAR model we follow a RL strategy and adjust the sequence length via the actor-critic framework, while keeping the order of the user-item interactions and optimizing the joint loss function in Equation \ref{eq:loss}.

\begin{figure}[t]
    \centering
    \includegraphics[scale=0.18]{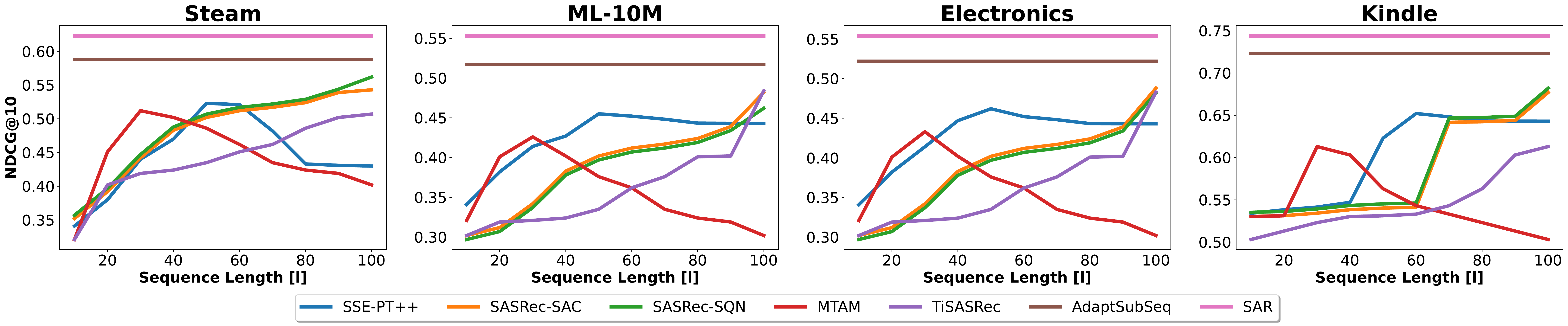}
    \caption{Impact of the sequence length $l$ on the recommendation performance of the examined models in terms of NDCG@10. The proposed SAR model and AdaptSubSeq follow adaptive sequence length strategies at every time step. Therefore, we report the same accuracy for the AdaptSubSeq and SAR models when varying the sequence length $l$.}
    \label{fig:ndcg_l}
\end{figure}

In Figure~\ref{fig:ndcg_l}, we present the impact of the sequence length $l$ on the performance of the examined models in terms of NDCG@10. Note that the proposed SAR model and AdaptSubSeq adjust the sequence length for each user at every time step. The average sequence lengths of the SAR model are $32.35 \pm 4.89$, $52.48 \pm 8.12$, $22.29 \pm 2.16$, and $67.92 \pm 14.22$ for the Steam, ML-10M, Electronics and Kindle datasets, respectively. In AdaptSubSeq, the average sequence lengths are $45.82 \pm 8.11$, $59.27 \pm 6.88$, $42.55 \pm 10.11$, and $88.91 \pm 6.23$, accordingly. On inspection of Figure~\ref{fig:ndcg_l}, we observe that SASRec-SAC, SASRec-SQN and TiSASRec require a significantly high sequence length to achieve a high recommendation accuracy, whereas SSE-PT++ and MTAM achieve the best accuracy when the sequence length is $50$ and $30$, respectively. Figure~\ref{fig:ndcg_l} clearly shows the importance of the sequence adaptation strategy of the proposed SAR model, as the baseline strategies underperform for all the different sequence lengths. Note that the baseline strategies with fixed sequence lengths might introduce noise to the recommendation accuracy for each user by either omitting recent sequential interactions or adding redundant interactions, which explains their limited recommendation accuracy when compared with the proposed SAR model.

\section{Conclusions} \label{sec:conc}

In this study we presented the SAR model, a sequence adaption model that follows a deep RL strategy. For each time step, the proposed SAR model adapts the sequence length in a personalized manner by formulating a joint loss function, where the Q-values of the RL agent act as a regularization to the sequential recommendation model. In doing so, the sequential recommendation model is aligned with the reward of the agent and SAR adjusts the sequence length in a personalized manner at the same time. Our experimental evaluation on four real-world datasets showed that SAR constantly outperforms the baseline approaches, achieving relative improvements of $6.4$ and $15.1\%$ in terms of NDCG and HR, respectively, when compared with the second best approach. An interesting future direction is to explore the continual RL approaches in the sequential recommendation problem~\cite{kaplanis2020continual,pmlr-v97-kaplanis19a}. The main challenge is not only to provide a lifelong learning model that continuously optimizes the policy of the agent but also to learn how to update the neural architecture over the sequential user-item interactions.

% \begin{figure}
%     \centering
%     \begin{array}{cc}
%         \includegraphics[scale=0.3]{figures/Steam.pdf} & \includegraphics[scale=0.3]{figures/ML10M.pdf} \\  \includegraphics[scale=0.3]{figures/Electronics.pdf} & \includegraphics[scale=0.3]{figures/Kindle.pdf} 
%     \end{array}
%     \caption{Caption}
%     \label{fig:my_label}
% \end{figure}

\bibliographystyle{ACM-Reference-Format}
\bibliography{sample-base}


\end{document}